# Performance Analysis on Molecular Dynamics Simulation of Protein Using GROMACS

A.D.Astuti and A.B. Mutiara
Department of Informatics Engineering, Faculty of Industrial Technology
Gunadarma University
E-mail: amutiara@staff.gunadarma.ac.id

**Abstracts**

*Development of computer technology in chemistry, bring many application of chemistry. Not only the application to visualize the structure of molecule but also to molecular dynamics simulation. One of them is Gromacs. Gromacs is an example of molecular dynamics application developed by Groningen University. This application is a non-commercial and able to work in the operating system Linux. The main ability of Gromacs is to perform molecular dynamics simulation and minimization energy.*

*In this paper, the author discusses about how to work Gromacs in molecular dynamics simulation of some protein. In the molecular dynamics simulation, Gromacs does not work alone. Gromacs interact with pymol and Grace. Pymol is an application to visualize molecule structure and Grace is an application in Linux to display graphs. Both applications will support analysis of molecular dynamics simulation.*

**Keywords:** molecular dynamics, Gromacs, Linux, pymol, grace, chemical

## 1. Introduction

### 1.1. Backgrounds

Computer is necessary for life of society, especially in chemistry. Now, many non-commercial application of chemistry is available in Windows version and also Linux. The applications are very useful not only in visualization molecule structure but also to molecular dynamics simulation.

Molecular dynamics is a simulation method with computer which allowed representing interaction molecules of atom in certain time period. Molecular dynamics technique is based on Newton law and classic mechanics law. Gromacs is one of application which able to do molecular dynamics simulation based on equation of Newton law. Gromacs was first introduced by Groningen University as molecular dynamics simulation machine.

### 1.2. Definition Of Problem

This writing is focused at usage of Gromacs application. In this writing, writer tell about how to install Gromacs, Gromacs concepts, file format in Gromacs, Program in Gromacs, and analysis result of simulation.

## 2. Theory

### 2.1. Protein

Protein is complex organic compound that has a high molecular weight. Protein is also a polymer of amino acid that has been linked to one another with a peptide bond.

Structure of protein divided into three, namely the structure of primary, secondary, tertiary and quaternary. Primary structure is amino acid sequence of a protein linked to it through a peptide bond. Secondary structure is a three-dimensional structure of local range of amino acids in a protein stabilized by hydrogen bond. Tertiary structure is a combination of different secondary structures that produce three-dimensional form. Tertiary structure is usually a lump.

Some of the protein molecule can interact physically without covalent bonds to form a stable oligomer (e.g. dimer, trimer, or kuartomer) and form a Quaternary structure (e.g. rubisco and insulin).

## 2.2. Molecular Dynamics

Molecular dynamics is a method to investigate exploring structure of solid, liquid, and gas. Generally, molecular dynamics use equation of Newton law and classical mechanics.

Molecular dynamics was first introduced by Alder and Wainwright in the late 1950s, this method is used to study the interaction hard spheres. From these studies, they learn about behavior of simple liquids. In 1964, Rahman did the first simulations using realistic potential for liquid argon. And in 1974, Rahman and Stillinger performed the first molecular dynamics simulations using a realistic system that is simulation of liquid water. The first protein simulations appeared in 1977 with the simulation of the bovine pancreatic trypsin inhibitor (BPTI) [8].

The main purposes of the molecular dynamics simulation is:

- Generate trajectory molecules in the limited time period.
- Become the bridge between theory and experiments.
- Allow the chemist to make simulation that can't bo done in the laboratory

## 2.3. The Concepts of Molecular Dynamics

In molecular dynamics, force between molecules is calculated explicitly and the motion of is computed with integration method. This method is used to solve equation of Newton in the constituents atomic. The starting condition is the position and velocities of atoms. Based on Newton's perception, from starting position, it is possible to calculate the next position and velocities of atoms at a small time interval and force in the new position. This can be repeated many times, even up to hundreds of times.

Molecular dynamics procedure can be described with the flowchart as follows:

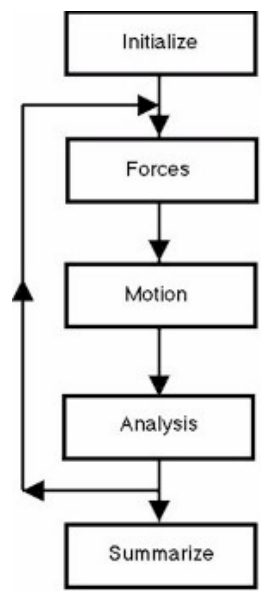

**Figure 2.1** Flowchart of molecular dynamics [13]

From The figure above can be seen the process of molecular dynamics simulation. The arrow indicates a path sequence the process will be done. The main process is calculating forces, computing motion of atoms, and showing statistical analysis the configuration for each atom.

## 3. Discussion

### 3.1. Gromacs Concepts

Gromacs is an application that was first developed by department of chemistry in Groningen University. This application is used to perform molecular dynamics simulations and energy minimization.

The concept used in Gromacs ia a periodic boundary condition and group. Periodic boundary condition is classical way used in Gromacs to reduce edge effect in system. The atom will be placed in a box, surrounded by a copy of the atom.

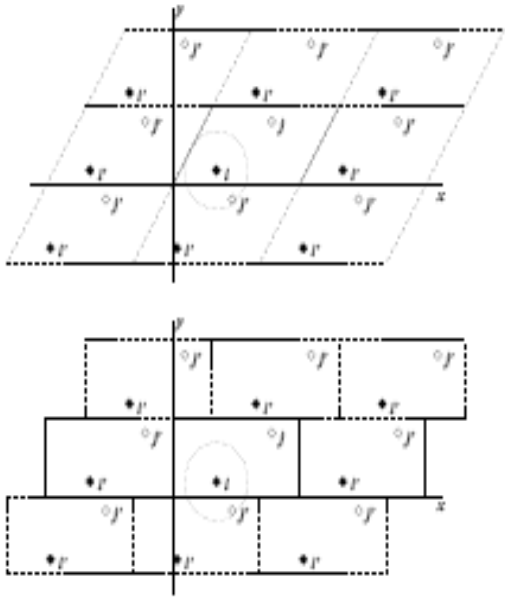

**Figure 2.2**: Periodic boundary condition in Two Dimensions [7]

In Gromacs there are some model boxes. That is triclinic, cubic, and octahedron.

The second concept is group. This concept is used in Gromacs to show an action. Each group can only have a maximum number of 256 atoms, where each atom can only have six different groups.

### 3.2. Install Gromacs

Gromacs applications can run on the operating system Linux and windows. To run Gromacs on multiple computer, then the required MPI (Message Passing Interface) *library* for parallel communication.

Gromacs applications can be downloaded in http://www.gromacs.org. How to install Gromacs is as follows:

1. Download FFTW in http://www.fftw.org
2. Extract file FFTW

    **% tar xzf fftw3-3.0.1.tar.gz**

    **% cd fftw3-3.0.1**

3. Configuration

    **% ./configure --prefix=/home/anas/fftw3 --enable-float**

4. Compile fftw

    **% make**

5. Installing fftw

    **% make install**

6. After fftw installed then install Gromacs. Extract Gromacs.

    **% Tar xzf gromacs-3.3.1.tar.gz**

    **% cd gromacs-3.3.1**

7. Configuration

    **% Export CPPFLAGS =-I/home/anas/fftw3/include**

    **% export LDFLAGS=-L/home/anas/fftw3/lib % Export LDFLAGS =- L/home/anas/fftw3/lib**

    **%. /configure – prefix=/home/anas/gromacs %. / Configure-prefix = / home / Anas / gromacs**

8. Compile and install gromacs

    **%make & make install**

### 3.3. Flowchart Of Gromacs

Gromacs need several steps to set up a file input in the simulation. The steps can be seen in flowchart below:

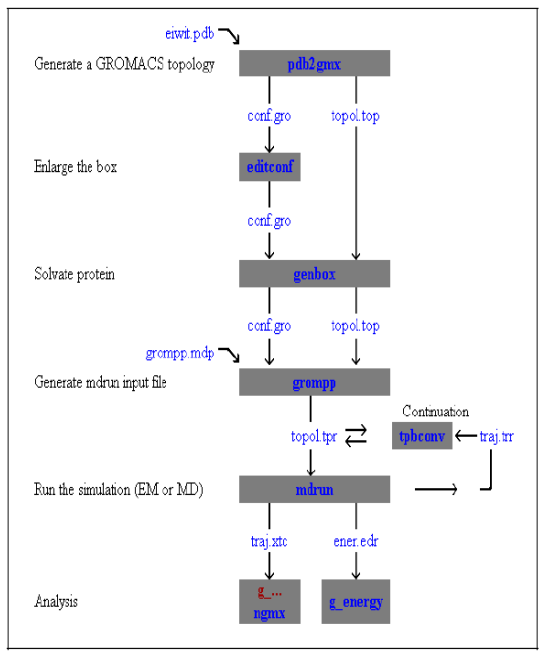

**Figure 3.1** Flowchart Gromacs [16].

Flowchart above illustrates how to do molecular dynamics simulation of a protein. The steps are divided into:

1. Conversion of the pdb file

    At this step pdb is converted to gromos file (gro) with pdb2gmx. Pdbgmx also created topology file (.top)

2. Generate box

    At this step, the editconf will determine the type of box and the box size that will be used in the simulation. on Gromacs there are three types of box, namely triclinic, cubic, and octahedron.

3. Solvate protein

    The next step is solvate the protein in box. The program genbox will do it. Genbox will generate a box defined by editconf based on the type. Genbox also determined the type of water model that will be used and add number of water molecule for solvate protein the water model commonly used is SPC (Simple Point Charge).

4. Energy minimization

    The process of adding hydrogen bond or termination may cause atoms in protein too close, so that the collision occurred between the atoms. The collision between atoms can be removed by energy minimization. Gromacs use mdp file for setup parameters. Mdp file specified number of step and cut-off distance. Use grompp to generate input file and mdrun to run energi minimization. The energy minimization may take some time, depending on the CPU [21].

5. Molecular dynamics simulation

    The process of molecular dynamics simulation is the same as energy minimization. Grompp prepare the input file to run mdrun. Molecular dynamics simulations also need mdp file for setup parameters. Most option of mdrun on molecular dynamics is used in energy minimization except –x to generate trajectory file.

6. Analysis

    After the simulation has finished, the last step is to analyze the simulation result with the following program:

    - Ngmx to perform trajectory
    - G_energy to monitor energy
    - G_rms to calculated RMSD (root mean square deviation)

### 3.4. File Format

In Gromacs, there are several types of file format:

1. Trr: a file format that contains data trajectory for simulation. It stores information about the coordinates, velocities, force, and energy.

2. Edr: a file format that stores information about energies during the simulation and energy minimization.
3. Pdb: a form of file format used by Brookhaven protein data bank. This file contains information about position of atoms in structure of molecules and coordinates based on ATOM and HETATM records.
4. Xvg: a form of file format that can be run by Grace. This file is used to perform data in graphs.
5. Xtc: portable format for trajectory. This file shows the trajectory data in Cartesian coordinates.
6. Gro: a file format that provides information about the molecular structure in format gromos87. The information displayed in columns, from left to right.
7. Tpr: a binary file that is used as input file in the simulation. This file can not be read through the normal editor.
8. Mdp: a file format that allows the user to setup the parameters in simulation or energy minimization.

### 3.5. Gromacs Programs

#### 3.5.1. Pdb2gmx

Pdb2gmx is a program that is used to convert pdb file. Pdb2gmx can do some things such as reading file pdb, adding hydrogen to molecule structure, and generate coordinate file a topologi file.

#### 3.5.2. Editconf

Editconf is used to define box water that will be used for simulation. This program not only defines the model, but also set the relative distance between edge of box and molecules. There are 3 types of box such as

- Triclinic, a box-shaped triclinic
- Cubic, a square-shaped box with all four side equal
- Octahedron, a combination of octahedron and dodecahedron.

#### 3.5.3. Grompp

Grompp is a pre-processor program. Grompp have some ability that is:

- Reading a molecular topology file
- Check the validity of file.
- Expands topology from the molecular information into the atomic information.
- Recognize and read topology file (*. top), the parameter file (*. tpr) and the coordinates file (*. gro).
- Generate *. tpr file as input in the molecular dynamics and energy of contraction that will be done by mdrun.

Grompp copy any information that required on topology file.

#### 3.5.4. Genbox

Genbox can do 3 things:

- Generate solvent box
- Solvate protein
- Adding extra molecules on random position

Genbox remove atom if distance between solvent and solute is less then sum of vanderwalls radii of each atom.

#### 3.5.5. Mdrun

Mdrun is main program for computing chemistry. Not only performs molecular dynamics simulation, but it can also perform Brownian dynamics, Langevin dynamics, and energy minimization. Mdrun can read tpr as input file and generate three type of file such as trajectory file, structure file, and energy file.

## 4. Result Of Simulation

### 4.1. Research Method

The testing carried out on different types of protein. Each protein has different structure and number of atom. Testing is based on flowchart of gromacs. This testing do two process, the first is energy minimization and the second is molecular dynamics simulation. Number of step for energy minimization is 200 numstep and molecular dynamics is 500 numstep.

### 4.2. Analysis of Simulation

From the testing, that was made on 4 different types of protein that can be seen the difference form of molecule before and after simulation. In molecular dynamics simulation, occur changes mechanisms of protein structure from folded state to unfolded state. Its mechanism is as follows:

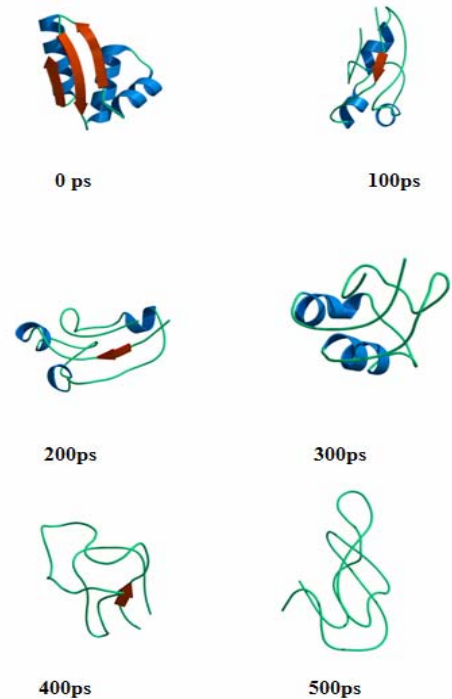

**Figure 4.1** Mechanism of Unfolded State [16]

In the molecular dynamics simulation above, each protein has a different velocity simulation.

**Table 4.1** Simulation Time Table

| Protein | Number of Atom | Simulasi ( minute: sec) for 500 iterations | |
|---|---|---|---|
| | | Minute:sec | sec |
| Alpha-Lactalbulmin | 7960 | 34:07 | 2047 |
| 1gg1-kappa d1.3 fv (Light Chain) | 2779 | 20:07 | 1207 |
| Ribonuleoside-Diphosphate Reductase 2 Alpha | 5447 | 3:30 | 210 |
| Lysozyme C | 1006 | 1:02 | 62 |

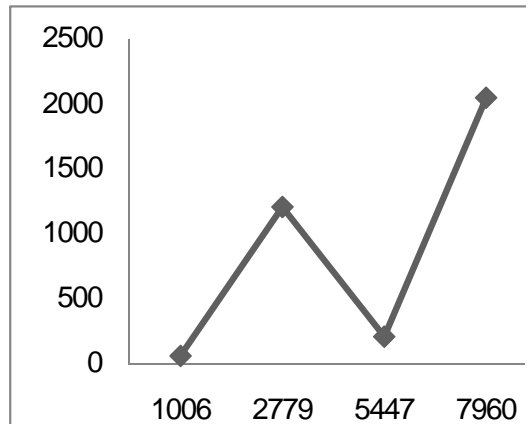
**Figure 4.2** Time Graphic Simulation

From the data above show differences long simulations of each protein. Length of time the simulation is depicted with a non-linier graph.

Length of time simulation is not only influenced by the number of atoms but also the number of chain and water blocks.

In the case of protein Ribonuleoside-Diphosphate Reductase Alpha 2, although the number of atom is greater than the protein 1gg1 FV-d1.3 Kappa (Light Chain) but the simulation time is more quickly. Because the number of blocks and the chain of water in this protein are lower than the protein 1gg1 FV-d1.3 Kappa (Light Chain).

## 5. Conclusion

Development of the world's computing allows the chemist to perform molecular dynamics simulation on hundreds of or thousands of atoms with computer. Molecular dynamics simulation is a technique for investigating structure of atoms based on the interaction with other atoms.

This writing introduces Gromacs as one of the applications that are able to perform molecular dynamics simulation, especially for protein. At this writing, the testing is carried out on four different types of protein. From The results of testing, it can be seen that each protein has a different long time.

At the protein Alpha-Lactalbulmin with number of atom 7960, long simulation time is 34 minutes 7 seconds. 1gg1 FV-d1.3 Kappa (light chain) with number of atom 2779, long simulation time is 20 minutes 7 seconds. Ribonuleoside-Diphosphate Reductase Alpha 2 with number of atom 5447, long simulation time is 3 minutes 30 seconds. And Lysozyme C with the number of atom 1006, long simulation time is 1 minute 2 seconds.

In addition gromacs also help understand the mechanisms Folding and unfolding of protein.